\documentstyle[aps]{revtex} 
\begin{document} 
\draft 
\title{New parity-violating photon-axion interaction}
\author{Parthasarathi Majumdar}
\address{The Institute of Mathematical Sciences, Chennai 600 113,
India\\Email:partha@imsc.ernet.in} 
\maketitle 
\begin{abstract} 
A variant of an earlier proposal by the author and SenGupta, to describe
four dimensional Maxwell electrodynamics in Einstein-Cartan spacetimes
through a Kalb-Ramond field as an intermediary, is shown to lead to a new
Maxwell-Kalb-Ramond coupling that violates spatial parity, even when
the KR gauge field has its standard parity assignment. One consequence
of this coupling seems to be a modulation, independent of
wavelength but dependent on the KR field strength, of the intensity of
synchrotron radiation observed from distant galactic sources.
\end{abstract}

There has been a resurgence of interest in cosmologies of spacetimes with
torsion and other generalizations of general relativity, in view of
precision experiments in cosmology including the recent launching of the
Microwave Anisotropy Probe and the forthcoming gravitational wave
detection experiments. A crucial issue in torsion-based cosmologies deals
with torsion coupling of all matter (and radiation) fields. For the
Maxwell field, the standard field strength tensor provides an adequate
means of description in Einstein-Cartan (EC) spacetimes, insofar as the
principle of general covariance is concerned \cite{obu}. Such a
description can be expanded to include a possible gauge invariant coupling
of the Maxwell field directly to the torsion. This particular issue has
been addressed, for the special case of a completely antisymmetric torsion
\cite{ms}, through the introduction of an antisymmetric second rank tensor
gauge potential $B_{\mu \nu}$ also known as the Kalb Ramond (KR) field. 

To effect a gauge invariant torsion coupling, the KR field strength
$H_{\mu \nu \lambda} \equiv \partial_{[\mu} B_{\nu \lambda]}$ needs to be
augmented by a $U(1)$ Chern-Simons form, in accord with certain
consistency requirements of quantum heterotic string theory \cite{gsw}. 
Now, in string theory, especially in its non-perturbative formulations
including the so-called Dirichlet branes \cite{pol}, there exist tensor
gauge fields of rank higher than 1, with rank 2 and 3 gauge fields being
the only non-trivial possibilities in four spacetime dimensions. In this
paper, we study possible augmentations akin to that proposed in \cite{ms},
to investigate torsion coupling of these higher rank gauge fields,
motivated primarily by an attempt to describe these fields by means of a
uniform framework. However, as we shall see shortly, gauge invariant
torsion couplng of these higher rank fields requires a modification of our
earlier procedure. It is not clear yet if this modification is essential
in the underlying string theory. However, if we insist on such a
modification also in the case of the Maxwell field, a new gauge invariant
Maxwell-KR interaction emerges that violates spatial parity, {\it
irrespective} of the parity of the KR field. This interaction has definite
observable consequences which we briefly describe, postponing a fuller
discussion to a later paper. 

For reasons of completeness, we begin with a brief review of \cite{ms}. 
The `stringy' augmentation of the KR field strength is given by
\begin{equation}
H_{\mu \nu \lambda} \rightarrow {\tilde H}_{\mu \nu \lambda} ~ \equiv~
H_{\mu \nu \lambda} ~+~ \frac13 \sqrt{G} A_{[\mu} F_{\nu \lambda]}
~,\label{htil}
\end{equation}
where, $G$ is Newton's constant. This augmented KR field strength is then
made to couple to torsion by  means of a simple contact interaction, so
that the ECKR-Maxwell action is
\begin{eqnarray}
S ~&=&~ \int d^{4}x~ \sqrt{-g}~[~{1 \over 16\pi G}~ R~(g~,~T)~-~\frac14
F_{\mu \nu}
~F^{\mu
\nu}~-\frac12~{\tilde H}_{\mu \nu \lambda} {\tilde H}^{\mu \nu
\lambda}~\nonumber \\
&+&~{1 \over \sqrt{G}}~T^{\mu \nu \lambda}~{\tilde H}_{\mu \nu \lambda}~]~
\label{act}   
\end{eqnarray}
where $R$ is the scalar curvature. The torsion tensor $T_{\mu \nu
\lambda}$ now plays the role of an auxiliary field in eq. (\ref{act}),
obeying the constraint equation   
\begin{equation}
T_{\mu \nu \lambda}~~=~~\sqrt{G}~{\tilde H}_{\mu \nu \lambda}~.
\label{tors}
\end{equation}
Thus, the augmented KR field strength three tensor plays the role of the
spin angular momentum density which is the source of torsion
\cite{Hehl}. In fact, (\ref{tors}) can be used to eliminate the torsion
from the theory and yield an `on-shell' action in which the Maxwell-KR
interaction, to leading order in the Planck length $\sqrt{G}$ is given by
\begin{equation}
S_{int}~=~\sqrt{G}~\int ~d^4x~\sqrt{-g}~H_{\mu \nu \rho} A^{[\mu} F^{\nu
\rho]}
~.\label{int}
\end{equation}
Now, the KR three tensor is Hodge-dual to the derivative of a spinless
field
$H$, so that, after a partial integration, one obtains,
\begin{equation}
S_{int}~=~\frac12~\sqrt{G}~\int d^4 x ~H~F_{\mu \nu}~^*F^{\mu
\nu}~,\label{inter}
\end{equation}  
where, $^*F^{\mu \nu}~\equiv~\epsilon^{\mu \nu \lambda \sigma}F_{\lambda
\sigma}$. 

We may mention that proposals for interactions depicted in eq. (\ref{int})
have appeared in earlier literature (see, e.g., references cited in
\cite{ms}) as well as in papers by Sikivie and collaborators \cite{sik}.
However, these proposals did not emerge from any microscopic consideration
based on a fundamental theory like string theory. Our attempt in \cite{ms}
has been to point out this microscopic underpinning, thus linking possibly
observable phenomena with fundamental physics. Moreover, the string
theoretic basis provides precise coupling strengths necessitated by
quantum consistency in that theory, thus cutting down on possible
theoretical uncertainties. Note also that the motivation for the
augmentation of the KR field strength came from the requirement of a gauge
invariant {\it direct} non-minimal coupling of the Maxwell field to
torsion.

The consequences of this interaction between the Maxwell and KR fields
have been explored \cite{kmss} within the context of synchrotron radiation
from cosmologically distant radio galaxies and quasars (see also
\cite{sik} for earlier phenomenological work in flat spacetime). The most
important effect appears to be an optical activity in the form of a
rotation of the plane of polarization of the radiation through an angle
that is proportional to the time rate of change of the axion field that is
Hodge-dual to the KR field strength \cite{kmss}. This rotation is
independent of wavelength for small wavelengths, and in that is to be
contrasted with Faraday rotation. Another feature of this optical activity
is its `universality' in that it is quite independent of source properties
(unlike, for instance, the Faraday rotation measure), although it does
depend on the redshift of the source. In the string-inspired situation,
the KR field $B_{\mu \nu}$ is an even spatial parity tensor so that the
spinless $H$ field is a {\it pseudo}scalar - the axion. It is obvious in
this case that the interaction in (\ref{inter}) preserves spatial parity.
However, one can consider the case where $B_{\mu \nu}$ has odd parity so
that $H$ is an even parity scalar; the interaction (\ref{inter}) in that
case violates spatial parity. This latter situation is interesting from
the point of view of polarization anisotropies of the Cosmic Microwave
Background Radiation (CMBR): certain odd-parity multipole moments of the
polarization tensor turn out to be non-vanishing if such an interaction is
present \cite{lue}. It is also interesting for its implications in high
energy physics \cite{bs} involving, e.g., helicity-flip scattering of
Dirac fermions \cite{ss}.

It is not difficult to see that a naive application of the `minimal
coupling' procedure (to
an EC spacetime) has the same pathology with gauge invariance for second
rank ($A_{\mu \nu}$) and third rank (i.e., $A_{\mu \nu \rho}$) Abelian
antisymmetric tensor gauge fields as for the Maxwell field
$A_{\mu}$. E.g., let us take the second rank antisymmetric gauge field
(a different KR field, considered as a gauge field rather than a source
of torsion). If one defines the spacetime covariant field strength as 
\begin{equation}
F_{\mu \nu \rho} ~\equiv~D_{[\mu}~A_{\nu \rho]}~ \label{kr2}
\end{equation}
one observes that under the tensor gauge transformation
\begin{equation}
\delta_{\Lambda}~A_{\mu \nu}~=~D_{[\mu} \Lambda_{\nu]}~,
\label{gauge} \end{equation}
the field strength transforms as
\begin{equation}
\delta_{\Lambda}~F_{\mu \nu \rho}~=~T_{[\mu
\nu}^{~\eta}~[\partial_{\eta}
\Lambda_{|\rho]} - \partial_{\rho]} \Lambda_{\eta}] ~+~`TT'~, \label{crv}
\end{equation}
where, $`TT'$ signifies terms that are quadratic in the torsion. The $rhs$
of (\ref{crv}) is certainly non-vanishing and signifies a violation of
tensor gauge invariance.

One may try to adapt the procedure outlined earlier to deal with this
problem for rank 2 and rank 3 fields, by using an augmented KR field
\'a la \cite{ms}. The
problem is with the augmentation. Using differential forms, it is
immediately obvious that the Chern-Simons augmentation $\sqrt{G} A \wedge
F$ vanishes in four dimensions for gauge connection $p$-forms $A$ for
$p>1$, where $F$ is the curvature $p+1$-form corresponding to $A$. To
construct a 3-form which one can append on to the KR 3-form, one can try
instead 
\begin{equation}
{\tilde{\bf H}} ~=~{\bf H}~+~\frac13~\sqrt{G} A \wedge~ ^*F~\label{new}
\end{equation}
where $^*F$ is the Hodge-dual of the curvature $p+1$ form, and ${\bf H}$
is the KR 3-form (with coordinate-frame components $H_{\mu \nu \rho}$). 

With the KR 3-form thus augmented, the procedure of ref. \cite{ms}
outlined above can be followed to determine the coupling of the tensor
gauge fields to torsion and eventually to the KR field itself. One obtains
generic KR-tensor gauge field interactions, given schematically to
$O(\sqrt{G})$, by  
\begin{equation}
S_{int} \sim \sqrt{G} ~\int d^4x ~\sqrt{-g}~H~F \cdot F~, \label{pint}
\end{equation} 
i.e., a coupling of the axion $H$ to the kinetic energy part of the
Lagrangian of the tensor gauge field. It is obvious that with the axion
being a pseudoscalar, interaction terms like (\ref{pint}) {\it violate
spatial parity}. While it is not clear if augmentations as in
(\ref{new}) arise naturally within string theories, interactions as
above may have interesting observable consequences. 

A particularly engaging case in point is that of the Maxwell field.
Combining (\ref{htil}) and (\ref{new}), one can define an `improved'
augmented KR field strength 3-form as
\begin{equation}
{\tilde{\bf H}} ~=~{\bf H}~+~\frac13~\sqrt{G}~[ \alpha_-~A \wedge~
^*F~+~\alpha_+~A \wedge~F]~, \label{htild}
\end{equation}
where, $\alpha_{\pm}$ are dimensionless real numbers. Thus, one is lead to
the $O(\sqrt{G})$ interaction 
\begin{equation}
S_{int} \sim \sqrt{G} ~\int ~H~[\alpha_-~F \cdot
F~+~\alpha_+~F \cdot~ ^*F]. \label{pvio}
\end{equation}  
With $H$ the axion, as is the situation in heterotic string theory, for 
example, the first term in (\ref{pvio}) is parity-violating and
the second, parity-preserving. To emphasize, unlike in earlier work
\cite{bs}, \cite{ss}, where one needed the KR gauge potential to have the
`wrong' (i.e., odd) parity in order to produce parity-violating effects,
here it is instead an augmentation of the KR field strength suggested by
an attempt to generalize the procedure to the case of higher rank tensor
gauge fields, that is responsible for parity-violation. 

Because of the difference in the parity behaviour between the two terms in
(\ref{pvio}), the physical consequences stemming from them are quite
distinct, at least to leading order in the Planck scale. In other words,
the consequences of the parity-preserving term in terms of a cosmological
optical activity \cite{kmss} are by no means affected by the presence of
the parity-violating interaction.

To examine the consequences of the parity-violating interaction, we once
again consider synchrotron radiation from distant galactic sources. Here
we confine ourselves to probing the effect of this term on a plane
electromagnetic wave in a Minkowskian background spacetime. The more
realistic case of radiation from large redshift sources in an expanding
universe will be taken up elsewhere. Thus, in this paper, the idea is to
anticipate the kind of phenomenon that may occur in realistic situations
due to the new interaction. 

To this end, we consider the Maxwell equations to linear order in
$\sqrt{G}$, following ref. \cite{kmss}, and treat the KR field as a
`cosmic' background which has decoupled from cosmological evolution,
earlier than the decoupling of photons. The idea thus, as in ref.
\cite{kmss}, is to consider the effect of this cosmic KR background on the
polarization of synchrotron radiation from large redshift galaxies. In
other words, while the coupled KR-Maxwell equations imply that the Maxwell
field `backreacts' on the KR field, producing inhomogeneous terms of the
type $\sqrt{G}~F \cdot F$ or $\sqrt{G}~ F \cdot ^* F$ in the KR field
equations, these backreaction effects will contribute to the Maxwell
equations only to $O(G)$ and not to the leading $O(\sqrt{G})$. Our
interest in this letter is confined to these latter leading $O(\sqrt{G})
\sim O(M_{Planck}^{-1})$ effects; the $O(G) \sim O(M_{Planck}^{-2})$
corrections may be considered in later analyses. To this extent, it is
correct to ignore the backreaction effects of the KR field and take the
dynamics of the KR field to be basically that of the free theory. It
follows that the Hodge-dual axion field satisfies field equations emerging
from the Bianchi identity that the KR field strength must be subject to
due to its own gauge invariance. Thus, while the axion field in general
must satisfy an equation of the type $\Box H = O(\sqrt{G})$, the leading
$O(\sqrt{G})$ effects in the Maxwell equations stem from solutions of the
{\it free} massless Klein-Gordon equation $\Box H =0$. 

The Maxwell equations now assume the form
\begin{eqnarray}
{\bf \nabla} \cdot {\bf E}~~&=&~\sqrt{G}~{\bf \nabla}H \cdot
[\alpha_+~{\bf
B}~+~\alpha_-~{\bf E}] \nonumber \\
{\dot {\bf E}}~-~{\bf \nabla} \times {\bf B}~~&=&~-\sqrt{G}\{\alpha_+~  [
{\dot H} {\bf B}~-~{\bf \nabla}H \times {\bf E}]
\nonumber \\
&&~+~\alpha_-~[{\dot H} {\bf E}~-~{\bf \nabla}H \times {\bf B}] \}
\nonumber \\
{\dot {\bf B}}~+~{\bf \nabla} \times {\bf
E}~&=&~0~=~{\bf
\nabla} \cdot {\bf B} ~.~\label{max2}
\end{eqnarray}

With potential cosmological applications in mind, we consider the simplest
situation in which the axion is spatially independent, as was assumed in
\cite{kmss}. This implies that the axion field has a linear time
dependence $H(t) = h_0 t + h_1$, with the constants $h_0, h_1$
being odd under parity, in accord with the intrinsic parity of the
axion.  The $\nabla H$ terms in (\ref{max2}) can
now be dropped, and the resulting equations manipulated to eliminate the
${\bf E}$ field for example. The resulting `wave equation' for the ${\bf
B}$ field is 
\begin{equation}
{\ddot {\bf B}}~-~\nabla^2 {\bf B} ~+~ \sqrt{G}~h_0~[\alpha_+~ \nabla
\times {\bf B}~+~\alpha_-~{\dot {\bf B}} ]~=~0~ . \label{mag}
\end{equation}
To further simplify the notation, we rescale $h_0 \rightarrow
h_0/\sqrt{G}$ and choose the ansatz
\begin{equation}
{\bf B}~=~{\rm Re}~[{\bf b}(t)~e^{i {\bf k} \cdot {\bf x}} ]~. \label{bee}
\end{equation}
Defining components of right and left circular polarization $b_{\pm}
\equiv b_x \pm i b_y$, with $z$ as the direction of propagation,  
eq. (\ref{mag}) reduces to,
\begin{equation}
{\ddot b_{\pm}}~+~h_-~{\dot b}_{\pm}~+~\left (k^2 \mp h_+~k
\right) b_{\pm}~=~0~, \label{fin}
\end{equation} 
where, $h_{\pm} \equiv h_0~\alpha_{\pm}$. 
Observe that (\ref{fin}) reduces, for $h_-=\alpha_-=0$, to eq. (14) of
ref. \cite{kmss} which leads to the optical activity already
discussed. Here, the angle of rotation of the plane of polarization of the
electromagnetic wave is, for large values of $k$, once again given by
$|h_+|~t$. The effect of parity violation is confined to the second term,
which signifies a {\it modulation}, i.e., either an enhancement or an
attenuation, of the intensity of the observed radiation, depending on the
sign of $h_-$. 

There are a few features of the new modulation effect due to the
parity-violating interaction that should be commented upon.

First of all, the modulation is proportional to the time rate of change of
the axion field and not just to its absolute value, unlike the rotation
angle of the polarization plane discerned in \cite{kmss}. From a
cosmological standpoint, one expects it to be negative, because of
expansion effects of the universe. There is dependence also on the
undetermined coefficient $\alpha_-$ which occurs in the proposed
augmentation (\ref{htild}) of the KR field strength. If $\alpha_- >0$,
then with $h_-<0$, eq. (\ref{fin}) would appear to predict an {\it
enhancement} in the intensity of radiation - a rather startling result.
For $\alpha_- < 0$, in contrast, one would observe an attenuation of that
intensity. This uncertainty in the physical outcome appears to be
unavoidable within the somewhat phenomenological approach of this work. 
Recall that in the augmentation of the KR field proposed in \cite{ms},
such an uncertainty in the coupling does not exist, as the coefficient is
fixed essentially by consistency requirements in string theory. Here, as
already remarked, one is not certain that there is such a microscopic
underpinning of the proposed augmentation.

Secondly, the first derivative term in (\ref{fin}) and hence the
modulation is {\it independent} of the wavelength of the radiation. This
indicates a modulation of the intensity which is different,
for example, from the one that seems to emerge \cite{kms} when one
considers an axion field that is harmonic in its spatial dependence
(rather than being spatially-independent as considered here); the damping
term in that case is proportional to the ratio of the wavelength of the
axion to that of the radiation field.

Thirdly, it is independent of the intrinsic properties
of the galactic source emitting the radiation or indeed the intergalactic
medium it traverses and in that sense `cosmological' in character. Thus,
it is unlike viscous damping caused by scattering of the wave
during its propagation through galactic and intergalactic medium. In a
true cosmological scenario, we anticipate the modulation to depend on the
redshift of the source. 

Fourthly, if the KR field is a {\it pseudo}tensor of rank 2, the
Hodge-dual field $H$ is an even parity scalar field. In that case, there
is an immediate consequence of dramatic importance, namely a
time-variation of the fine structure constant. Recently, a lot of
attention has focussed on observation of such variation in the shift of
spectral lines seen from distant quasi-stellar sources \cite{webb}.
Note also that the second term in the action in (\ref{pvio}) now violates
spatial parity with consequences for the CMBR polarization anisotropies as
analyzed in \cite{lue}. Work is in progress to determine if these two
effects could be related within the framework discussed here.

Finally, the key outcome of the foregoing analysis is the inevitability of
spatial parity violation due to the interaction (\ref{pvio}) irrespective
of the parity of the KR field. It is not unlikely that the freedom to
carry out the proposed modified augmentation of the KR field strength
originates in the underlying string theory (at least in the zero slope
limit) from the freedom to add finite local counterterms to the effective
action without changing the anomaly cancellation conditions. Thus,
interactions like in (\ref{pvio}) could in fact be generic in string
theory. 

Note, {\it en passant} that, just as the Chern Simons augmentation has a
gravitational
analogue, given by the 3-form $\omega \wedge R$ where $\omega$ is the
Lorentz spin connection one-form and $R$ is the Riemann curvature
two-form, there ought to be `parity-violating' gravitational augmentations
characterised by the 3-form $\omega \wedge~ ^*R$ where $^*R$ is the dual
curvature. Thus, one anticipates new parity-violating gravitational
interactions of the type
\begin{equation}
S_{int} ~\sim \int ~ H~R \cdot R~.
\end{equation}
One effect that such an interaction might produce could be a modulation of
gravitational waves. If it turns out that the analogue of $\alpha_-$ in
that case is positive, while the time rate of change of the axion is
negative, one will be led to the truly remarkable prediction of
enhancement of the intensity of gravitational waves from distant sources.
This does seem to warrant a more thorough investigation.

\end{document}